\documentclass[a4paper]{jpconf}
\usepackage{graphicx}

\newcommand{\cp}{\ensuremath{c_{p}}}
\newcommand{\cel}{\ensuremath{c_{\rm el}}}
\newcommand{\cph}{\ensuremath{c_{\rm ph}}}

\newcommand{\gn}{\ensuremath{\gamma_{\rm n}}}
\newcommand{\gs}{\ensuremath{\gamma_{\rm s}}}
\newcommand{\gres}{\ensuremath{\gamma_{\rm res}}}

\newcommand{\TD}{\ensuremath{{\it \Theta}_{\rm D}}}
\newcommand{\Hc}{\ensuremath{H_{\rm c}}}

\newcommand{\Tc}{\ensuremath{T_{\rm c}}}

\begin{document}
\title{Specific heat of aluminium-doped superconducting silicon carbide}

\author{M~Kriener$^1$, T~Muranaka$^2$, Y~Kikuchi$^2$, J~Akimitsu$^2$ and Y~Maeno$^1$}

\address{$^1$Department of Physics, Graduate School of Science, Kyoto University, Kyoto 606-8502, Japan}
\address{$^2$Department of Physics and Mathematics, Aoyama-Gakuin University, Sagamihara, Kanagawa 229-8558, Japan}

\ead{mkriener@scphys.kyoto-u.ac.jp}

\begin{abstract}
The discoveries of superconductivity in heavily boron-doped diamond, silicon and silicon carbide renewed the interest in the ground states of charge-carrier doped wide-gap semiconductors. Recently, aluminium doping in silicon carbide successfully yielded a metallic phase from which at high aluminium concentrations superconductivity emerges. Here, we present a specific-heat study on superconducting aluminium-doped silicon carbide. We observe a clear jump anomaly at the superconducting transition temperature 1.5\,K indicating that aluminium-doped silicon carbide is a bulk superconductor. An analysis of the jump anomaly suggests BCS-like phonon-mediated superconductivity in this system. 
\end{abstract}

\section{Introduction}
In 2007, superconductivity at $\Tc\approx 1.45$\,K was discovered in heavily boron-doped silicon carbide (SiC:B) at doping concentrations of $n\sim 10^{21}$\,cm$^{-3}$ by means of resistivity and AC susceptibility measurements \cite{ren07a}. The upper critical field \Hc\ was estimated to approximately 120\,Oe, i.e., a rather small value. This and the finding of a broad in-field hysteresis in temperature- (field-) dependent AC susceptibility measurements between cooling (field down sweep) and warming (field up sweep), known as \textit{supercooling}, lead to the surprising conclusion that SiC:B is a type-I superconductor. This is in clear contrast to the reported type-II superconductivity in the parent compounds boron-doped diamond (C:B) and silicon (Si:B) \cite{ekimov04a,bustarret06a}. Another remarkable difference among these systems is the polytypism of silicon carbide (SiC). Many different crystal modifications with energetically slightly different ground states are reported in literature. The most common ones are 3C-SiC, 2H-, 4H- and 6H-SiC, and 15R-SiC. The number in front of C (\,=\,cubic unit cell), H (\,=\,hexagonal) or R (\,=\,rhombohedral) indicates the number of Si\,--\,C bilayers stacking in the conventional unit cell. Please note that all SiC polytypes break inversion symmetry whereas both Si and C possess an inversion symmetry centre. The sample used in the study published in Ref.\,\cite{ren07a} consists of major phase fractions of 3C- and 6H-SiC, both of which do participate in the superconductivity \cite{kriener08b,kriener09a,muranaka09a}. 
Subsequently, a specific-heat study demonstrated that SiC:B is a bulk superconductor, likely to be described in a phonon-mediated BCS-type scenario of the superconductivity with a significant ($\sim 50$\,\%) residual density of states due to nonsuperconducting parts of the sample used. The normal-state Sommerfeld parameter estimates to $\gn\approx 0.29$\,mJ/molK$^2$ and the jump height fits the BCS expectation $\Delta \cel/\gs\Tc=1.43$ \cite{kriener08a}.
This paper focuses on aluminium-doped silicon carbide (SiC:Al). The preparation was done in a similar way as for SiC:B \cite{ren07a,kriener08a}. The sample used in this study belongs to the same batch used in Ref.\,\cite{muranaka09a}, where the finding of superconductivity in SiC upon aluminium doping at $\Tc\approx 1.5$\,K is reported. It is a multiphase polycrystalline sample mainly consisting of cubic 3C-SiC. In addition, pure Si and Al was identified by x-ray diffraction, but there is no indication of other SiC polytypes in this sample. Its charge-carrier concentration was estimated to $0.7\cdot 10^{21}$\,cm$^{-3}$ from a Hall-effect measurement, somewhat smaller than that found for the SiC:B sample. Though the superconducting transition temperatures are almost the same for our SiC:B and SiC:Al samples, the latter one does not exhibit a clear in-field hysteresis in temperature- and field-dependent resistivity measurements \cite{muranaka09a} suggesting that aluminium-doped SiC might be a type-II superconductor. Compared to SiC:B, this also places SiC:Al closer to C:B in the theoretical phase diagram proposed in Ref.\,\cite{yanase09a}. However, there is a possibility that the type-I behaviour in SiC:Al is ''hidden'' away by disorder and hence the supercooling behaviour is diminished, see Refs.\,\cite{muranaka09a} and \cite{blase09a}.

\section{Specific heat: Analysis and Discussion}
The specific-heat data presented in this paper was taken by a relaxation-time method using a commercial system (Quantum Design, PPMS).
\begin{figure}
\begin{center}
\includegraphics[width=7cm]{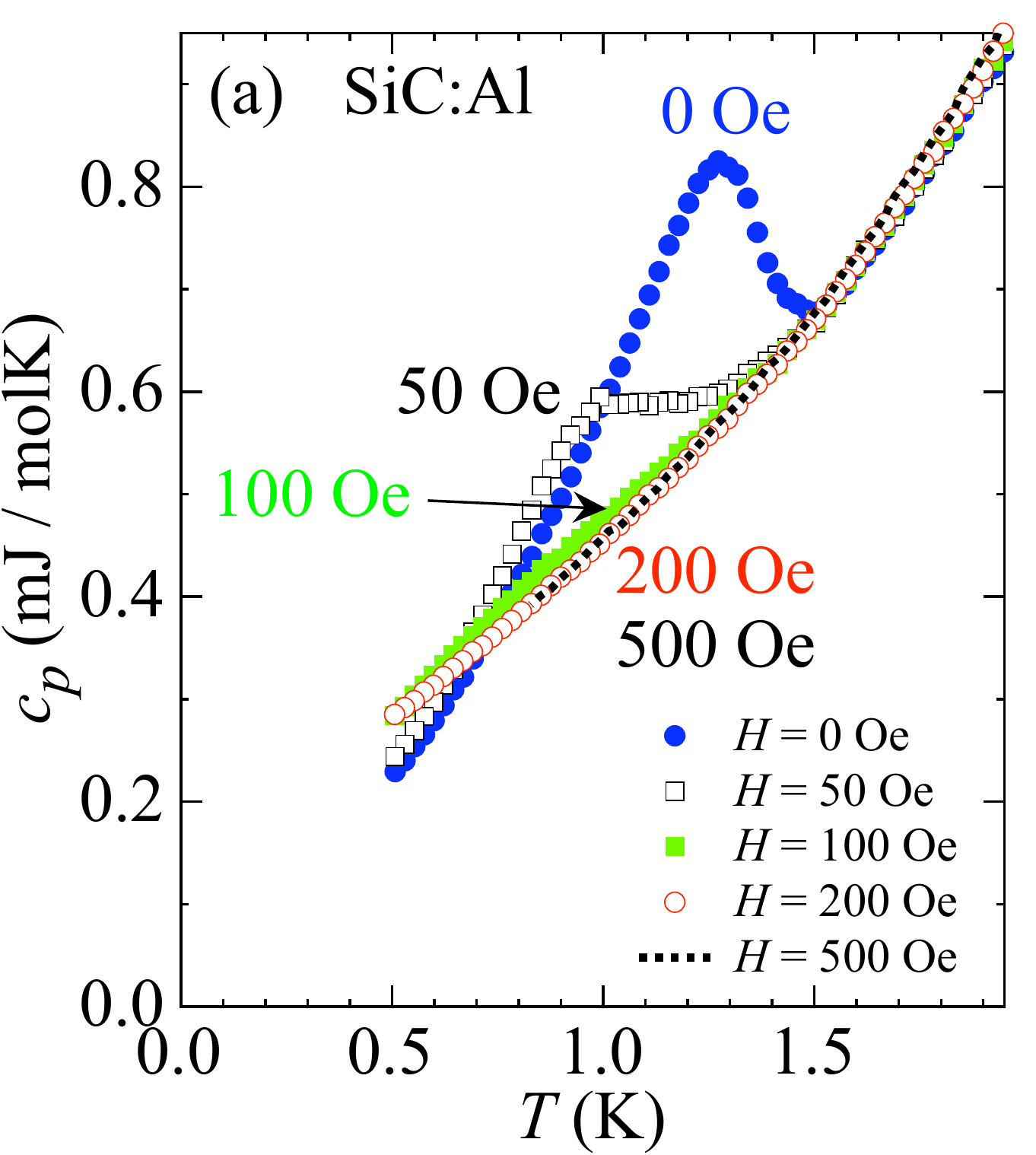}
\hspace{0.5cm}
\includegraphics[width=7cm]{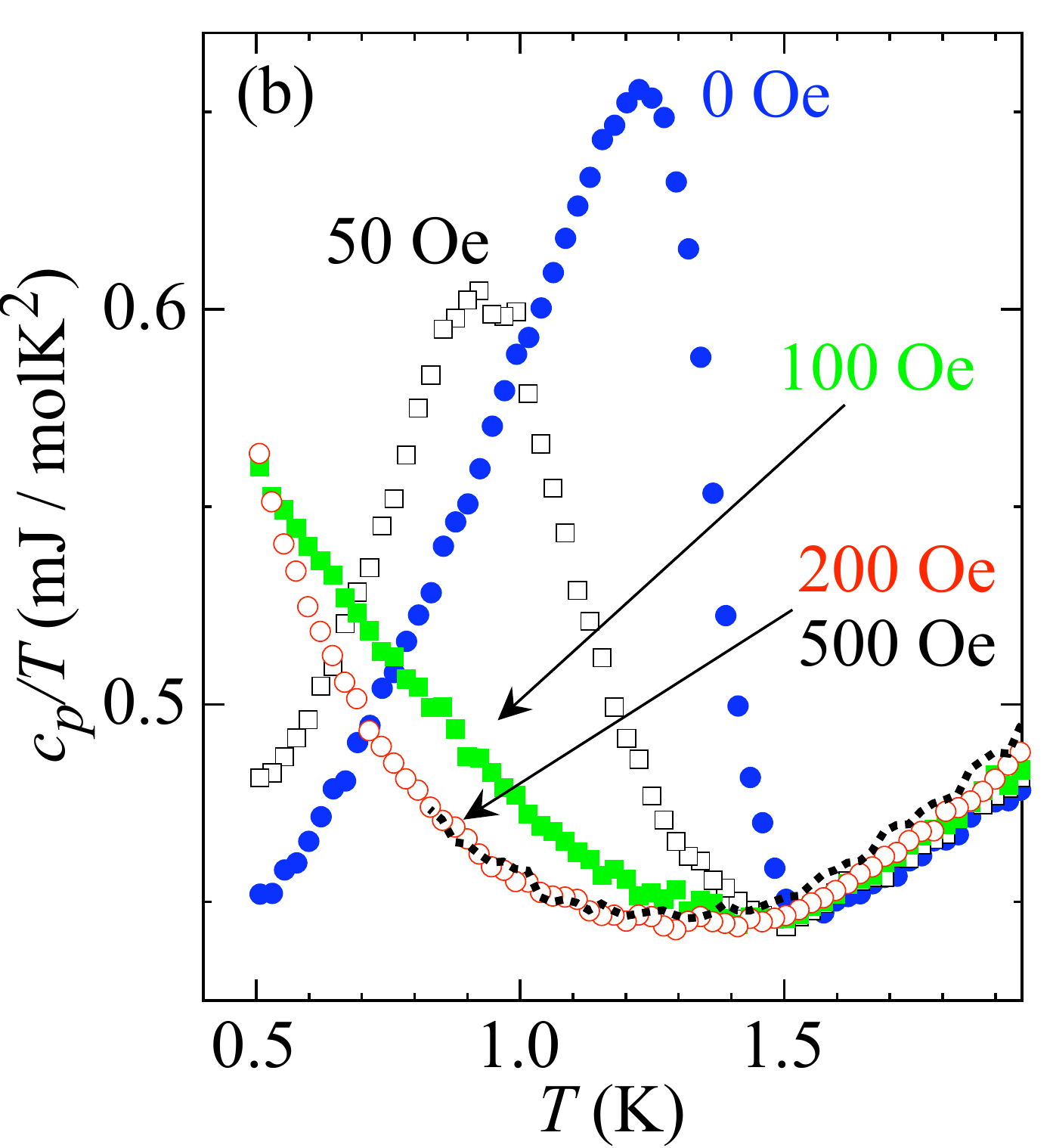}
\end{center}
\caption{\label{fig1} Specific heat of 3C-SiC:Al in $H=0$, 50, 100, 200 and 500\,Oe. In panel (a) the specific heat $\cp$ as measured is shown. Panel (b) gives the specific heat displayed as $\cp/T$ on an expanded scale. Note that both axes in panel (b) do not start at 0.}
\end{figure}
Figure\,\ref{fig1} summarises the temperature dependence of the specific-heat data of 3C-SiC:Al in various DC magnetic fields $H=0$, 50, 100, 200, and 500\,Oe. Panel (a) gives the specific heat \cp\ as measured and panel (b) gives an expanded view of the specific heat divided by temperature $\cp/T$. A clear jump anomaly in the zero-field data is observed at $\Tc\approx 1.5$\,K coinciding with the transition temperatures found in resistivity \cite{muranaka09a} and AC susceptibility measurements \cite{kriener09b}. Furthermore, the jump indicates that the superconductivity in this sample is a bulk feature. However, the transition is rather broad reflecting the polycrystalline multiphase character of the sample used. A field as small as 50\,Oe is sufficient to suppress the superconducting transition strongly and hence shifts \Tc\ to lower temperatures. In $H=100$\,Oe only a small anomaly is left. The data for $H=200$ and 500\,Oe are lying on top of each other, hence there is no indication of a bulk superconducting transition down to 0.5\,K in fields $H\geq 200$\,Oe. As seen in figure\,\ref{fig1}\,(b), there is an additional low-temperature contribution $c_{\rm add}$ which is not field dependent between 200\,Oe and 500\,Oe indicating that the same contribution is also present at lower fields but masked by the superconducting transition. The additional entropy responsible for this upturn in the data might arises from a nuclear Schottky effect due to the aluminium atoms doped into the system (nuclear spin of Al: $I=5/2$; carbon and silicon cannot be responsible for it due to their $I\leq 1/2$). This issue needs clarification, e.\,g., by extending the data to lower temperatures and higher fields, which is currently underway. 

\begin{figure}
\begin{center}
\includegraphics[width=14cm]{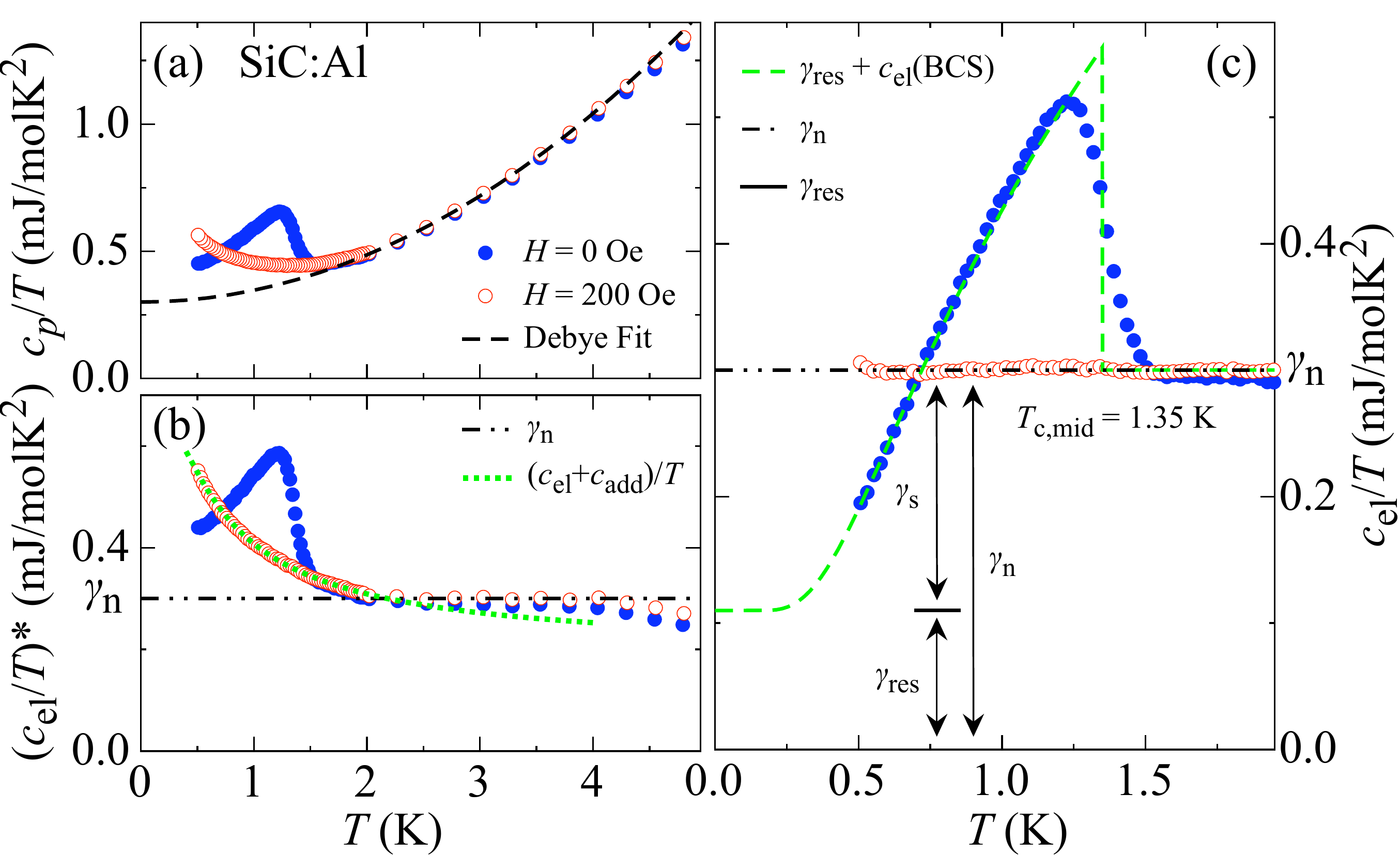}
\end{center}
\caption{\label{fig2} In panel (a) the specific-heat data $\cp/T$ in $H=0$\,Oe (closed symbols) and 200\,Oe (open symbols) from Fig.\,\ref{fig1} are shown again. The dashed line is a Debye fit to the data. Panel (b) gives the specific heat $(\cel/T)^*=(\cel+c_{\rm add})/T$ after subtracting the phononic contribution. The dashed-dotted line represents the normal-state Sommerfeld parameter, the dotted curve is a fit to the data $(\cel/T)^*$ to estimate the additional contribution $c_{\rm add}$ at low temperatures, see text. Subtracting this additional specific heat yields the electronic specific heat $\cel/T$ given in panel (c) where the dashed curve is a fit to the zero-field data assuming a BCS-like superconducting ground state in this system, see text.}
\end{figure}
In figure\,\ref{fig2}\,(a) the specific heat data $\cp/T$ in $H=0$\,Oe (closed symbols) and 200\,Oe (open symbols) are shown again. The latter field exceeds the (upper) critical field and hence represents the normal-state specific heat of 3C-SiC:Al. A fit to these in-field data in the temperature interval $2.4\,{\rm K} < T < 4$\,K applying the conventional Debye formula $\cp=\cph+\cel = \gn T+\beta T^3$ yields the normal-state Sommerfeld parameter $\gn=0.30$\,mJ/molK$^2$ and the prefactor of the phononic contribution to the specific heat $\beta = 0.05$\,mJ/molK$^4$. The corresponding curve is plotted dashed in figure\,\ref{fig2}\,(a). The Debye temperature evaluates to $\TD\approx 440$\,K which is much smaller than that for undoped SiC, i.e., 1200\,--\,1300\,K depending on the polytype; compare the discussion in Ref.\,\cite{kriener08a}. Subtracting the phononic contribution from the data shown in figure\,\ref{fig2}\,(a) yields the data $(\cel/T)^*=(\cel+c_{\rm add})/T$ displayed in figure\,\ref{fig2}\,(b). Therein the value of the normal-state Sommerfeld coefficient is indicated by the dashed-dotted line. To further analyse the data, it is necessary to subtract the additional contribution $c_{\rm add}$ to the specific heat below $2\sim 3$\,K leading to the aforementioned strong upturn in the data. Therefore the in-field data below 2\,K was fitted applying $(\cel/T)^*= aT^{-2}+bT^{-1}+c$ yielding the dotted curve in figure\,\ref{fig2}\,(b). Figure\,\ref{fig2}\,(c) gives the electronic specific heat $\cel/T$ after subtracting $c_{\rm add}$ assuming that the additional contribution is the same in both the zero-field and in-field data.
The resulting electronic specific-heat data in the superconducting state can be modeled assuming a BCS-type behaviour of $\cel/T$ below \Tc\ 
\begin{equation}\label{GlBCS_res}
\cel(T)/T = \gres + \gs/\gn\cdot \cel^{\rm BCS}(T)/T,
\end{equation}
as described in detail in Ref.\,\cite{kriener08a}.
Due to the polycrystalline character of the sample used, it is reasonable to assume that a certain fraction of the sample remains normal conducting allowing for a residual density of states contributing to the specific heat but not participating in the superconductivity. Therefore, a residual Sommerfeld coefficient $\gres=\gn-\gs$ is included in (\ref{GlBCS_res}) with \gs\ representing the superconducting part of the sample. In this approach, \gres\ is the only adjustable parameter. A fit to the zero-field data applying (\ref{GlBCS_res}) yields the dashed curve shown in figure\,\ref{fig2}\,(c) with $\gres\approx 0.11$\,mJ/molK$^2$ as indicated by arrows. This corresponds to a superconducting volume fraction of about 60\,\%. Entropy conservation at the transition yields the mid-point transition temperature $T_{\rm c, mid}=1.35$\,K, slightly lower than the onset temperature $\Tc=1.5$\,K onbtained from resistivity and AC susceptibility measurements. 

\section{Summary}
The present study reveals that aside from boron-doping also aluminium doping of silicon carbide leads to a metallic phase from which eventually bulk superconductivity emerges ($\Tc\approx 1.5$\,K) which can be explained in a BCS-type phonon-mediated scenario. The superconducting volume fraction of the sample used estimates to $\sim 60$\,\%.  Surprisingly, in spite of the different dopands, i.e., the different size of the doped atoms, these results on SiC:Al do not much differ, neither qualitatively nor quantitatively, from those reported for SiC:B. Therefore, superconductivity in the charge-carrier doped semiconductor silicon carbide remains a highly interesting topic. 

The origin of the strong enhancement of the specific heat below $\sim 2$\,K needs to be clarified. A nuclear Schottky effect due to the doped aluminium atoms might be responsible for the additional entropy at low temperatures.

\section*{Acknowledgments} 
We acknowledge technical support of C.\,Michioka. This work was supported by a Grants-in-Aid for the Global COE ''The Next Generation of Physics, Spun from Universality and Emergence'' from the Ministry of Education, Culture, Sports, Science, and Technology (MEXT) of Japan, and by the 21st century COE program ''High-Tech Research Center'' Project for Private Universities: matching fund subsidy from MEXT. It has also been supported by Grants-in-Aid for Scientific Research from MEXT and from the Japan Society for the Promotion of Science (JSPS). TM is supported by Grant-in-Aid for Young Scientists (B) (No. 20740202) from MEXT and MK is financially supported as a JSPS Postdoctoral Research Fellow. 

\section*{References}
\providecommand{\newblock}{}

\end{document}